\documentclass[12pt,preprint]{aastex}

\shorttitle{Twisted tubes}
\shortauthors{Zaqarashvili et al.}

\begin{document}

\title{Twisted magnetic flux tubes in the solar wind}

\author{Teimuraz V. Zaqarashvili\altaffilmark{1,3},
Zolt\'an V\"or\"os\altaffilmark{1}, Yasuhito Narita\altaffilmark{1}, and Roberto Bruno\altaffilmark{2}}

\altaffiltext{1}{Space Research Institute, Austrian Academy of Sciences,
Schmiedlstrasse 6, 8042 Graz, Austria. Email: teimuraz.zaqarashvili@oeaw.ac.at}
\altaffiltext{2}{INAF-IAPS Istituto di Astrofisica e Planetologia Spaziali, Rome, Italy} \altaffiltext{3}{Abastumani Astrophysical Observatory at
Ilia State University, Tbilisi, Georgia.}

\begin{abstract}
Magnetic flux tubes in the solar wind can be twisted as they are transported from the solar surface, where the tubes are twisted owing to photospheric motions. It is suggested that the twisted magnetic tubes can be detected as the variation of total (thermal+magnetic) pressure during their passage through observing satellite. We show that the total pressure of several observed twisted tubes resembles the theoretically expected profile. The twist of isolated magnetic tube may explain the observed abrupt changes of magnetic field direction at tube walls. We have also found some evidence that the flux tube walls can be associated with local heating of the plasma and elevated proton and electron temperatures. For the tubes aligned with the Parker spiral, the twist angle can be estimated from the change of magnetic field direction. Stability analysis of twisted tubes shows that the critical twist angle of the tube with a homogeneous twist is 70$^0$, but the angle can further decrease owing to the motion of the tube with regards to the solar wind stream. The tubes with a stronger twist are unstable to the kink instability, therefore they probably can not reach 1 AU.

\end{abstract}


\section{Introduction}\label{intro}

Turbulent fluctuations in the solar wind are increasingly dominated by magnetic energy at large heliospheric distances which can be partly associated with advected flux tubes \citep{bruno2007}. This idea is further supported by recent results about magnetic coherent structures (current sheets) which have been found to be locally associated not only with intermittency but also with temperature enhancements in the solar wind \citep{osman2012}. Current sheets in the solar wind can arise due to nonlinear turbulent interactions \citep{chang2004, servidio2009}, steepening of outward propagating Alfven waves \citep{malara1996} or can be flux tube walls \citep{borovsky2008}. It is crucial to understand the contribution of all these physical processes to intermittency or heating of solar wind plasma. In particular, dynamical evolution of flux tubes in the solar wind may be important for better understanding of turbulence and heating.

Each magnetic flux tube may contain a distinct plasma and may lead to the distinct feature of MHD turbulence. If the magnetic flux tubes are "fossil structures" (i.e. they are carried from the solar atmosphere), then they may keep the magnetic topology typical for tubes near the solar surface. The solar magnetic field has a complicated topology in the whole solar atmosphere. Photospheric motions may stretch and twist anchored magnetic field, which may lead to the consequent changes of topology at higher regions. The observed rotation of sunspots \citep{brown2003,zhang2007} may lead to the twisting of the magnetic tubes above active regions, which can be observed in chromospheric and coronal spectral lines \citep{srivastava2010}. Recent observations of magnetic tornados \citep{wedemeyer2012,li2012} also strongly support the existence of twisted magnetic flux tubes on the Sun. On the other hand, the newly emerged  magnetic tubes in the solar photosphere are supposed to be twisted during the rising phase through the convection zone \citep{moreno-insertis1996,archontis2004}. Consequently, magnetic flux tubes in the solar wind could be also twisted if they are originated in the solar atmosphere. The twisted tubes can be unstable to Kelvin-Helmholtz instability when they move with regards to the solar wind stream \citep{zaqarashvili2013}. The Kelvin-Helmholtz vortices may eventually lead to enhanced turbulence and plasma heating, therefore the twisted tubes may significantly contribute into solar wind turbulence.

Twisted magnetic tubes can be observed by in situ vector magnetic field measurements in the solar wind \citep{moldwin2000,feng2007,cartwright2010}, but the method establishes limitations on real magnetic field structure considering force-free field model. In this Letter, we suppose a new method for in situ observation of twisted magnetic flux tubes and study their stability assuming that the solar wind plasma is composed of individual magnetic flux tubes which are carried from the solar surface by the wind \citep{bruno2001, borovsky2008} or locally generated \citep{telloni2012}. We show that the twisted magnetic tubes can be observed in situ through observation of total (thermal+magnetic) pressure and this method removes the necessity of the force-free field consideration. In the next section, we calculate the total pressure of simple model twisted tube and compare it to real observations.

\section{Observation of twisted tubes through total pressure variation}

We consider isolated twisted magnetic flux tube of radius $a$ with the magnetic field of $(0,B_{\phi}(r),B_z(r))$ and the thermal pressure of $P_0(r)$ in the cylindrical coordinate system $(r,\phi,z)$, where $r$ is the distance from the tube axis. Integration of pressure balance condition inside the tube gives the expression for the total pressure as
\begin{equation}
\label{tot} P_{T0}(r)=P_{T0}(0)-{{1}\over {4 \pi}}\int^{r}_{0}{{{B^2_{\phi}(s)}\over {s}}ds},
\end{equation}
which shows that the total pressure $P_{T0}=P_0+ {{(B^2_{\phi}+ B^2_z)}/{8\pi}}$
is not constant inside the twisted magnetic flux tubes and its radial structure depends on the type of twist.

Figure 1 shows the radial structure of normalized total pressure (red solid line) inside the tube with simplest homogeneous twist of $B_{\phi}=Ar$, where $A$ is a constant. It is seen that the total pressure is maximal at the tube axis and decreases towards the surface. Therefore, the variation of total pressure observed in situ by satellite may indicate the passage of an isolated twisted tube. Untwisted tube yields no variation in the total pressure (see red dashed line on Fig.1), therefore it is easy to identify the passage of twisted and untwisted tubes. The type of twist can be deduced from Eq. (\ref{tot}) using the observed profile of total pressure.
In order to test the method, we consider already known observations of twisted magnetic flux tubes in the solar wind. WIND spacecraft magnetic field \citep{lepping1995} and plasma data \citep{lin1995,ogilvie1995} are used. For demonstration purposes, Fig. 2 shows an event in detail, which was first spotted by \citet{feng2007} and carefully analyzed  by \citet{telloni2012}. It was found that this event was embedded between regions of different solar wind speed. The analysis of magnetic helicity and hodogram revealed a twisted structure with a clockwise rotation of the magnetic field within the structure \citep{telloni2012}.
The z-component of the magnetic field in GSE system also shows recognizable bipolar turning (Figure 2c), which according to \citet{moldwin2000} is the classic flux rope signature (note that here $B_z$ is in the GSE system).
The walls of the flux tube structure indicated by vertical boxes before 22:30 UT on March 28 (leading edge), 1998 and at 02:15 UT on March 29, 1998 (trailing edge) are readily recognizable as jumps in the values of physical quantities. At the trailing edge a sudden increase of $T_p$ and a slight decrease of $T_e$ can be observed with  steady values afterwards. The temperature change, the increase of density and the single peaked jump in $\Delta \Theta$ indicate that the probe crosses the flux tube discontinuity and enters to a different plasma. Contrarily, at the leading edge, the density before and after the wall is the same, one can see a simultaneous increase of both $T_p$ and $T_e$ and several rapid fluctuations in the orientation of magnetic field vector ($\Delta \Theta$). Since before and after the leading edge (first vertical box in Figure 2) the temperatures are smaller and $\Delta \Theta$ fluctuates less, we interpret these observations as signatures of local heating at the unstable flux tube wall. We plot the total pressure ($P_T$) (Figure 2d), which is the sum of electron+proton thermal pressure and magnetic pressure. The enhanced $P_T$ at the center of the interval indicates a twisted magnetic field structure. Converting the temporal observations of the advected structures to spatial observations using the solar wind speed, we can compare our data with the theoretical total pressure profile for twisted flux tubes. The observed smoothed profile of total pressure (black solid line) is plotted over the theoretical profile of twisted tube (see Figure 1). It is seen that the theoretical curve describes the observed profile rather well. This means that the observed flux tube may have a homogeneous twist of $B_{\phi}=Ar$. The double peaked structure, which is mainly resembling the evolution of $B$ (Figure 2c,d), can appear as a result of periodic kinking motions of the flux tube or internal inhomogeneities via turbulence or waves. In consequence of the kinking motion $P_T$ can gradually increase, then decrease, before the subsequent increase near the center of the flux tube. Since the changes are rather smooth in time with a quasi-period $>$ 1 h (Figure 2d) we can suppose that the resulting double peaked structure is associated with kink motions rather than multi-scale turbulence. The other curves of total pressure in Figure 1 correspond to WIND flux tube events, listed in \citet{moldwin2000}, for which both proton and electron data
are available. These plots indicate that the total pressure has radial structure inside the flux ropes of \citet{moldwin2000}, therefore it is in agreement with our predictions. However, some flux tubes in \citet{moldwin2000} are less twisted than the event in Figure 2.

The twist angle near the tube surface for the model pressure profile (red solid line on Fig. 1) is estimated as $\sim$ 55$^0$. Then the flux tube of \citet{feng2007} and some tubes of \citet{moldwin2000} are twisted with the angle of 50$^0$-55$^0$, while the other tubes of \citet{moldwin2000} are less twisted.

The abrupt change of magnetic field direction $\Delta \Theta$ may indicate a crossing the wall of twisted tube if the axis of an isolated tube is aligned with the Parker spiral (Figure 3). Then the angle of abrupt change of magnetic field direction may show the twist angle at the tube wall, which could be significantly scattered from the direction of the Parker spiral. Indeed, in a statistical study, \citet{borovsky2008} found that the tube axes are aligned with the Parker spiral with significant scatter (Borovsky considered only untwisted tubes, therefore the direction of magnetic field inside the tube was considered as the tube axis). The observational scatter of tube magnetic field average direction with regards to the Parker spiral can be explained by the simultaneous existence of untwisted and twisted magnetic tubes in the solar wind plasma: untwisted tubes in average are aligned with the Parker spiral and the scatter is caused by the twisted tubes. However the hypothesis is too simplified and some spread in results is expected due to the complexity of solar wind plasma. \citet{borovsky2008} also found that the mean angle between the tube magnetic field and the Parker spiral is $\sim$ 42.7$^0$, while the mean angle between the wall normal and the Parker spiral is peaked towards 90$^0$. If the magnetic tube axes are aligned with the Parker spiral, then the result of \citet{borovsky2008} means that the mean twist angle of tubes is $\sim$ 42.7$^0$. On the other hand, the twisted tubes are subjects to the kink instability when the twist exceeds a critical value. Therefore it is important to estimate whether the angle is less than the critical one.
\section{Stability of twisted magnetic flux tubes}

Normal mode analysis \citep{dungey1954} and energy consideration method \citep{lundquist1951} show the similar thresholds of the kink instability in twisted magnetic tubes. The instability condition for the homogeneous twist $B_{\phi}=Ar$ and $B_z=const$ is $B_{\phi}(a)>2B_z$. This leads to the critical twist angle of $\sim$ 65$^0$. Therefore, the mean twist angle of $\sim$ 42.7$^0$ indicates that the majority of tubes are stable to the kink instability. External magnetic field may increase the threshold and thus stabilize the instability \citep{bennett1999}. On the other hand, a flow along the twisted magnetic tube may decrease the threshold \citep{zaqarashvili2010}. Magnetic flux tubes in the solar wind may move with regards to the main stream of solar wind particles, therefore it is important to study the competitive effects of external magnetic field and the motion of tube (or external medium). Note that the consideration is simplified compared to turbulent solar wind plasma.

In order to study the instability criterion of moving twisted magnetic flux tube in the external magnetized medium, we use the normal mode analysis. We consider a tube with the homogeneous twist, $B_{\phi}=Ar$, homogeneous axial magnetic field $B_{z}$ and uniform density $\rho_0$. The external medium with homogeneous magnetic field $(0,0,B_e)$ directed along the $z$-axis and the uniform density $\rho_e$ is moving with homogenous speed $U$ along the tube axis i.e. along the $z$-axis. It is equivalent to the consideration of moving magnetic tube with the speed of $-U$ in static external medium. In order to obtain the dispersion equation governing the dynamics of the tube, one should find solutions of perturbations inside and outside the tube and then merge them at the tube surface through boundary conditions. After Fourier analysis of linearized magnetohydrodynamic equations with $\exp[i(m\phi+kz-\omega t)]$ , where $k$ is longitudinal wavenumber and $\omega$ is the frequency, incompressible perturbations of total pressure, $p_t$, inside the tube are governed by the Bessel equation \citep{dungey1954,bennett1999,zaqarashvili2010,zhelyazkov2012}
\begin{equation}
\label{bessel} {{d^2p_t}\over {dr^2}}+{1\over r}{{d p_t}\over {dr}}-\left ({{m^2}\over {r^2}}+m^2_0\right )p_t=0,
\end{equation}
where
$$
m^2_0=k^2\left (1-{{4A^2\omega^2_A}\over {4 \pi \rho_0(\omega^2-\omega^2_A)^2}}\right ),\,\,\omega_A={{mA+kB_{z}}\over {\sqrt{4 \pi \rho_0}}}.
$$

The bounded solution of the equation is the modified Bessel function $p_t=a_0I_m(m_0 r)$, where $a_0$ is a constant. The perturbations outside the tube is governed by the same Bessel equation, but $m_0$ is replaced by $k$. The solution outside the tube bounded at infinity is $p_t=a_eK_m(k r)$, where $a_e$ is a constant. The boundary conditions at the tube surface are the continuity of Lagrangian displacement $[\xi_r]_a=0$ and total Lagrangian pressure $[p_t-B^2_{\phi} \xi_r/(4 \pi r)]_a=0$ \citep{dungey1954,bennett1999,zaqarashvili2010}, which after straightforward calculations give the transcendental dispersion equation
$$
{{(\omega^2-\omega^2_A)F_m(m_0a)-2m\omega_A A /\sqrt{4 \pi \rho_0}}\over {(\omega^2-\omega^2_A)^2-4\omega^2_A A^2/(4 \pi \rho_0)}}=
$$
\begin{equation}
\label{dispersion}={{P_m(ka)}\over {([\omega-kU]^2-\omega^2_{Ae})(\rho_e/\rho_0)+P_m(ka)A^2/(4 \pi \rho_0)}},
\end{equation}
where
$$
F_m(m_0 a)=´{{m_0aI^{'}_m(m_0a)}\over {I_m(m_0a)}}, \,\,P_m(ka)={{kaK^{'}_m(ka)}\over {K_m(ka)}},
$$
and $\omega_{Ae}=kB_{e}/\sqrt{4 \pi \rho_e}$. A prime ($'$) denotes the derivative of a Bessel function to its dimensionless
argument. Imaginary part of $\omega$ in the dispersion equation (Eq. \ref{dispersion}) indicates the instability of the tube. The threshold for the kink instability ($m=1$) can be found analytically through the marginal stability analysis, i.e. considering $\omega=0$ \citep{chandrasekhar1961}. Using the thin flux tube approximation, $ka \ll 1$ (yielding $F_1(m_0 a)\approx 1+m^2_0a^2/4+...$ and $P_1(ka)\approx -1$), after some algebra we obtain the following criterion for the kink instability from Eq. (\ref{dispersion})
\begin{equation}
\label{instability} B_{\phi}(a)>2B_{z}\left (1+{{k B_{z}}\over {A}}\right )\sqrt{1-{\rho_e\over \rho_0}M^2_A+\mu^2} ,
\end{equation}
where $M_A=U\sqrt{4\pi \rho_0}/B_{z0}=U/V_{A0}$ is the Alfv\'en Mach number and $\mu=B_{e}/B_{z}$ (here $U$ is the relative speed of tube with regards to the mean solar wind stream and could be much less than the wind speed itself). For a static tube with a field-free environment the criterion leads to the Lundquist criterion (see also \citet{dungey1954}). For $B_{e}$=0 it leads to the instability condition of twisted tube moving in a field free environment \citep{zaqarashvili2010}. The critical twist angle for the kink instability can be approximated as
\begin{equation}
\label{angle} \theta_c=\arctan\left ({{B_{\phi}(a)}\over {B_{z}}}\right )\approx \arctan\left (2\sqrt{1-{\rho_e\over \rho_0}M^2_A+\mu^2}\right ).
\end{equation}
It is seen that the critical twist angle depends on the Alfv\'en Mach number, the ratio of external to internal axial magnetic field strength $\mu$ and the ratio of external to internal densities. The larger twist angle than the critical one leads to the kink instability. Figure 4 shows that the critical twist angle decreases when $\mu$ decreases and $M_A$ increases. More dense tubes are more stable. Maximum critical twist angle $\sim$ 70$^0$ occurs for static tubes $M_A=0$ and $\mu=1$. It means that the tubes twisted with the angle of $\geq$ 70$^0$ are always unstable to the kink instability.

The kink instability may lead to the magnetic reconnection, which may either destroy the tube \citep{feng2011} or remove additional twist from the tube keeping only stable configuration. Therefore, the tubes twisted with the angle of $\geq$ 70$^0$ probably can not reach the distance of 1 AU. The motion of tube with the Alfv\'en speed with regards to the solar wind stream may reduce the critical twist angle to 45$^0$ for $B_{z}>B_{e}$ and $\rho_e$=0.8 $\rho_0$. This is close to the statistically mean value of the angle between tube magnetic field and the Parker spiral obtained by \citet{borovsky2008}. Three from five tubes analyzed in this letter have the twist angle of 50$^0$-55$^0$, which is less than the critical angle for the kink instability of the tubes with $M_A<0.5$, but can be larger for tubes with $M_A>0.5$. Suppose a magnetic flux tube is twisted with sub-critical angle near the Sun. The Alfv\'en Mach number could be increased towards the Earth owing to the decrease of Alfv\'en speed. Consequently, initially stable flux tube may become unstable to the kink instability at some distance from the Sun. Two other tubes are below the threshold of kink instability for any $M_A$. 

\section{Discussion and Conclusions}

It is shown in this letter that the twisted magnetic tubes in the solar wind can be detected by in situ observations as variation of total pressure during the passage of the tubes through a satellite. The method allows us to obtain the radial structure of the twist and it can be used for any configuration of the magnetic field including non force-free field. We tested the method using several already known cases of observed twisted tubes, which shows that the total pressure in observed events resemble the theoretically expected profile. Therefore, the total pressure variation can be used to estimate the value of twist and its radial structure in the tubes embedded in the solar wind. The method can be also used to estimate the twist in coronal mass ejections (CME) in addition to Grad - Shafranov Reconstruction method \citep{mostl2009}.

We suggest that the twist of isolated magnetic tube may explain the observed abrupt changes of magnetic field direction at tube walls in the solar wind \citep{borovsky2008}. The mean statistical angle of the abrupt change, which was estimated by \citet{borovsky2008} as $\sim$ 42.7$^0$, can be considered as the mean twist angle of the magnetic flux tubes. Observed significant scatter of tube magnetic field average direction with regards to the Parker spiral can be explained by untwisted and twisted magnetic tubes: untwisted tubes are aligned with the Parker spiral and the scatter is caused by the twisted tubes.

Using stability analysis of twisted magnetic tubes, we obtain the theoretical criterion of kink instability, which shows that the maximal twist angle is $\sim$ 70$^0$ in the case of static tubes, while it decreases to 45$^0$ for the tubes moving with the Alfv\'en speed with regards to the solar wind. It may explain the observed mean statistical angle of 42.7$^0$, because the tubes twisted with a larger angle are unstable to the kink instability, therefore they probably can not reach 1 AU.
%

Tangential velocity discontinuity due to the motion of magnetic flux tubes with regards to the solar wind stream may lead to the Kelvin-Helmholtz
instability \citep{drazin1981}. A flow-aligned magnetic field may stabilize sub-Alfv\'enic flows \citep{chandrasekhar1961}. However, the twisted tubes can be unstable for any sub-Alfv\'enic motions if they move with an angle to the Parker spiral \citep{zaqarashvili2013}. Then the Kelvin-Helmholtz vortices may lead to the enhanced MHD turbulence and plasma heating near the walls of twisted magnetic tubes.

Statistical fraction of twisted tubes in the solar wind may correspond to the fraction of twisted tubes near the solar surface, which is not known because of observational constraints. Therefore, in situ observations of twisted tubes in the wind may allow us to estimate their percentage in the solar lower atmosphere.

In conclusion, twisted magnetic flux tubes could be essential components in the solar wind structure and they may play significant role in the turbulence and heating of the solar wind plasma.

{\bf Acknowledgements} The work was supported by EU collaborative project STORM - 313038. The work of TZ was also supported by FP7-PEOPLE-2010-IRSES-269299 project- SOLSPANET, by Shota Rustaveli Foundation grant DI/14/6-310/12 and by the Austrian "Fonds zur F\"{o}rderung der wissenschaftlichen Forschung" under project P26181-N27. The work of ZV was also supported by the Austrian "Fonds zur F\"{o}rderung der wissenschaftlichen Forschung" under project P24740-N27. We acknowledge WIND spacecraft flux-gate magnetometer data from the Magnetic Field Investigation, plasma data from the
3D Plasma Analyser and from the Solar Wind Experiment.

\clearpage

\begin{figure}
\epsscale{1}
\plotone{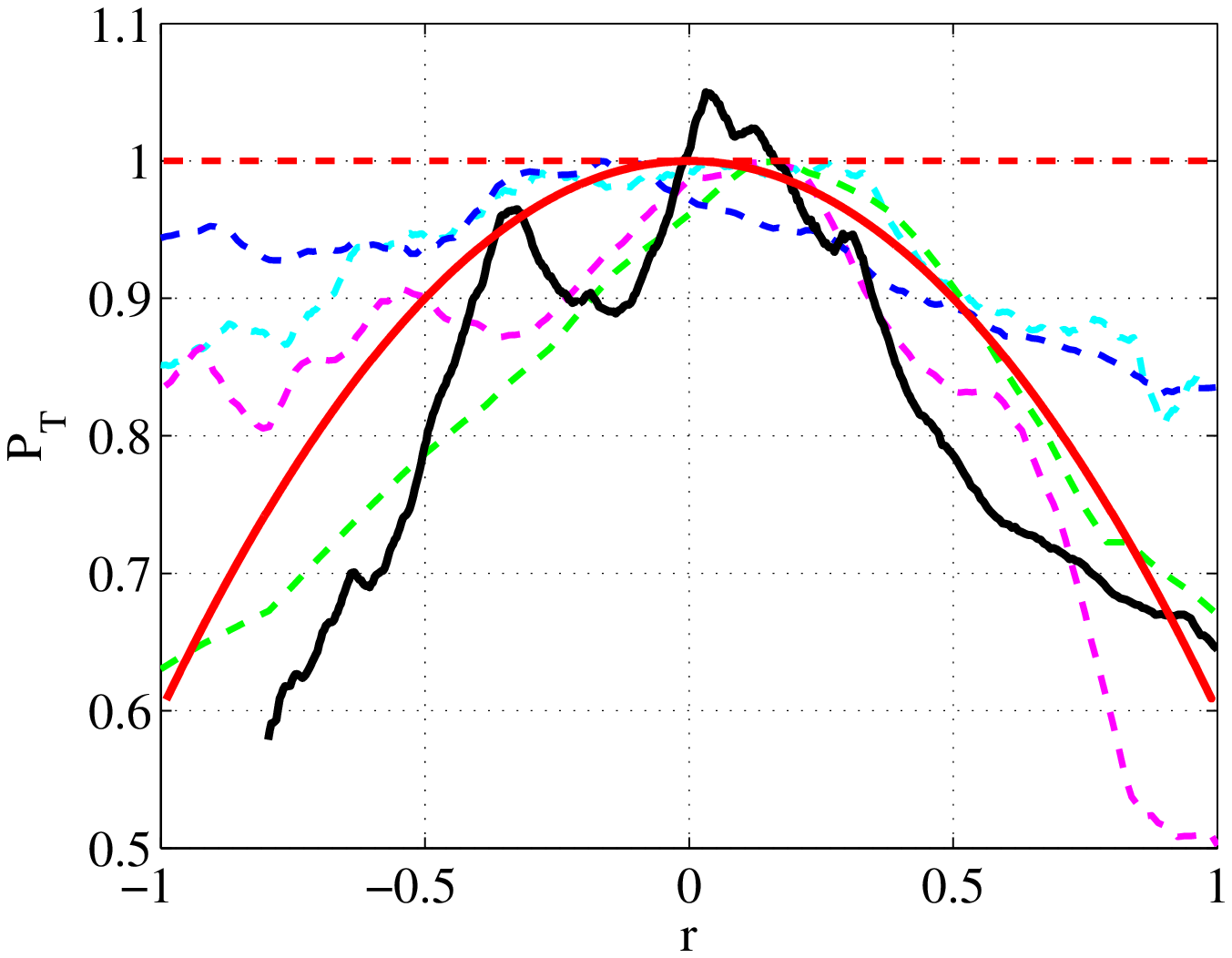}
\caption{The structure of total pressure inside model twisted (red solid line) and untwisted (red dashed straight line) magnetic flux tubes. The total pressure is normalized by its value at tube axis ($r=0$) and the distance from the tube axis is normalized by the tube radius $a$. Here the magnetic pressure of $\phi$ component at the tube surface is 0.4 of the total pressure at the tube axis.  Black solid line shows the radial structure of total pressure in magnetic flux tube first detected by \citet{feng2007} using WIND observations (see also Fig.2). Blue, green, purple and cyan dashed curves correspond to the total pressure profiles of flux ropes analyzed by \citet{moldwin2000}.} \label{fig1}
\end{figure}

\begin{figure}
\epsscale{1}
\plotone{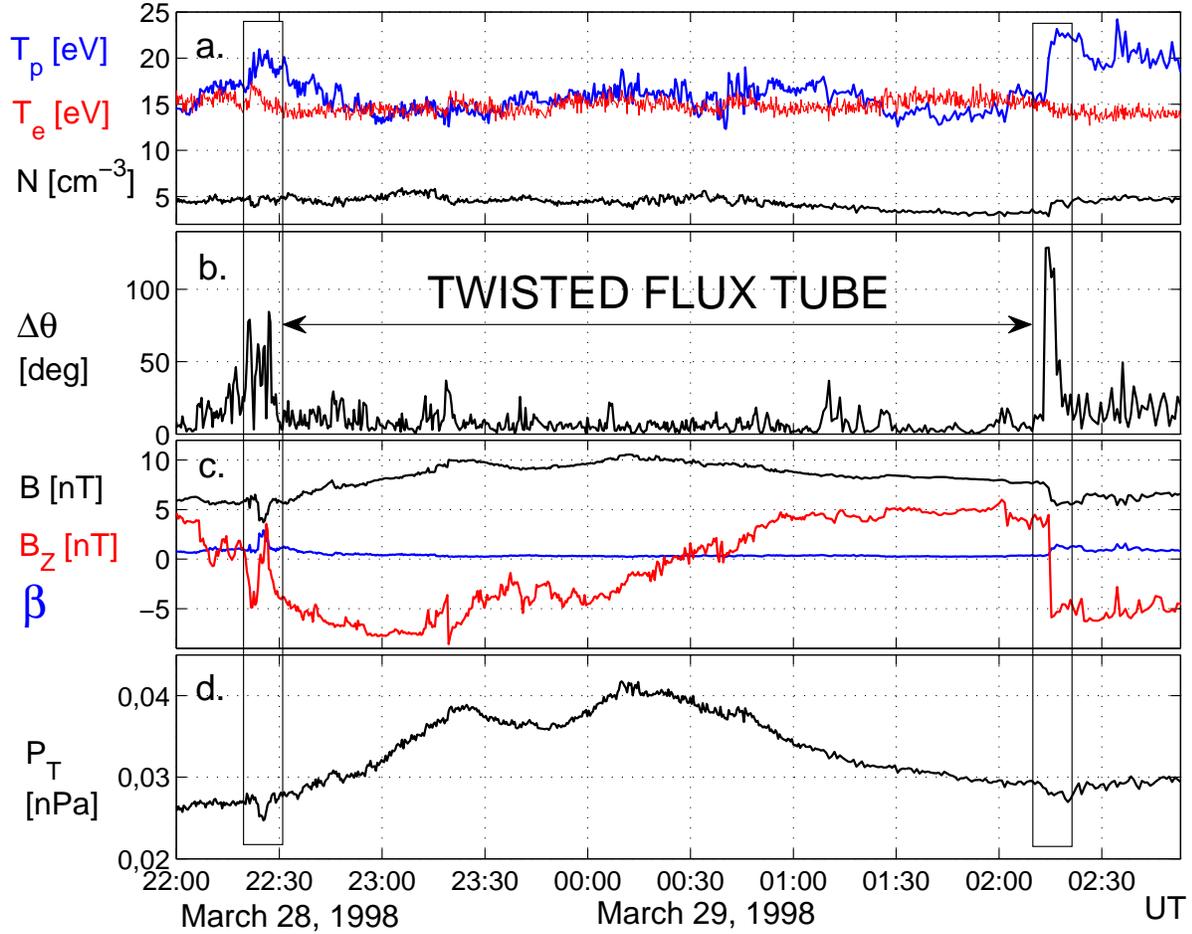}
\caption{WIND observations of a twisted flux tube: a. proton ($T_p$) temperature, electron ($T_e$) temperature and proton density ($N$); b. Magnetic field vector directional change $\Delta \Theta$ over the time scale of 2 minutes; c. Total magnetic field ($B$), the magnetic field z-component in GSE system ($B_z$) and plasma $\beta$; d. Total pressure ($P_T$) as a sum of magnetic, proton and electron pressures.}\label{fig2}
\end{figure}

\clearpage

\begin{figure}
\epsscale{1}
\plotone{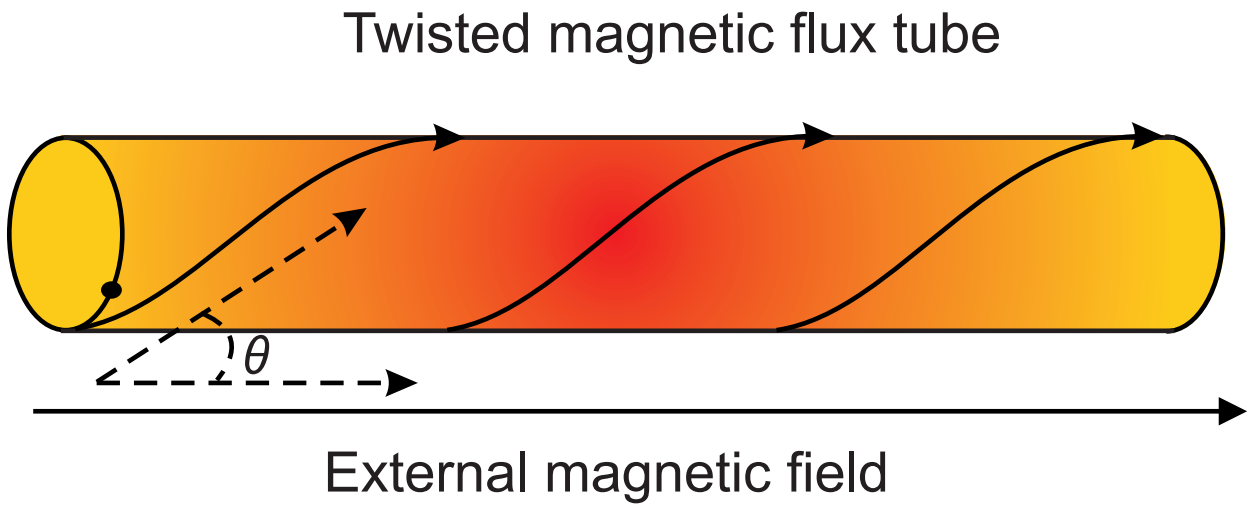}
\caption{Twisted magnetic flux tube in the solar wind. Tube axis is supposed to be directed along mean external magnetic field, which can be the Parker spiral. $\theta$ is the angle between the internal twisted magnetic field and the Parker spiral. Tangential discontinuity of magnetic field strength and direction can be detected when the tube passes through a satellite.}\label{fig3}
\end{figure}

\begin{figure}
\epsscale{1}
\plotone{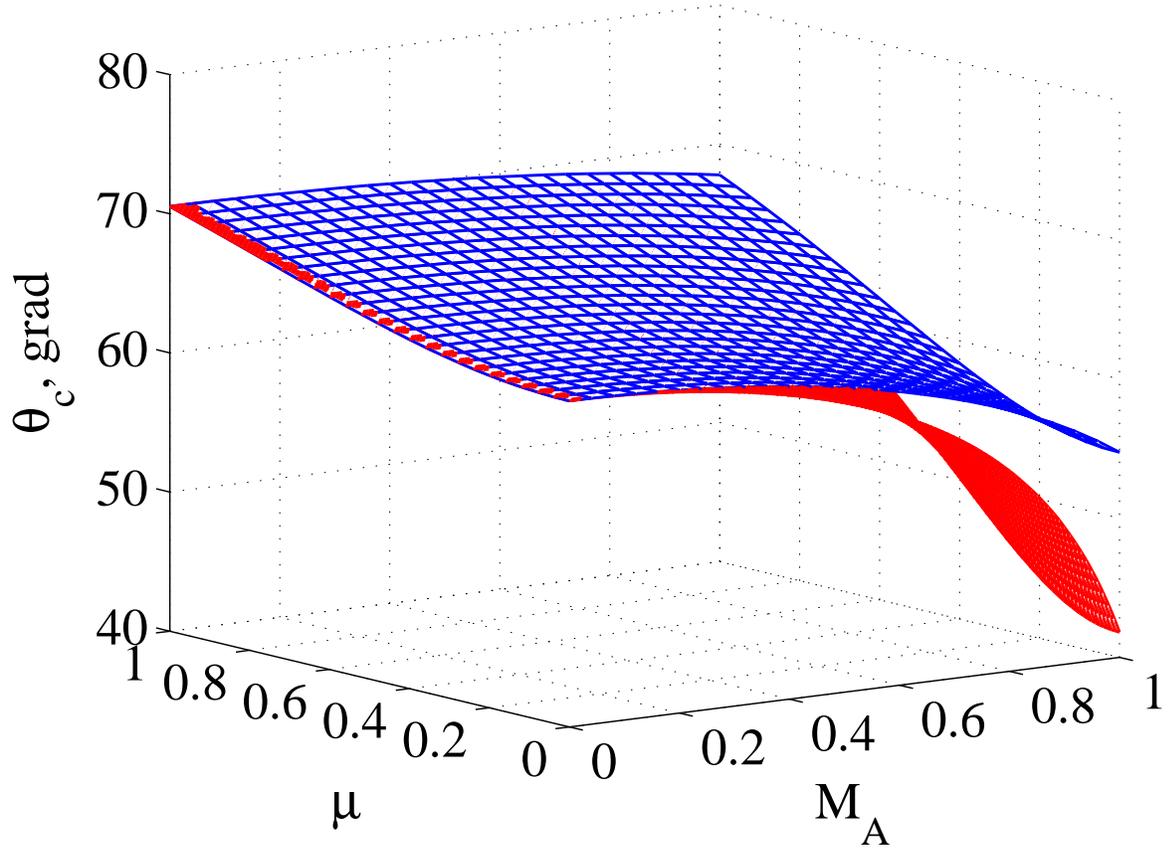}
\caption{Critical twist angle vs Alfv\'en Mach number, $M_A$, and the ratio of external and internal axial magnetic field strengths, $\mu=B_{e}/B_{z}$, as calculated from Eq. (\ref{angle}). Here the red (blue) surface corresponds to $\rho_e/\rho_0=0.8$ ($\rho_e/\rho_0=0.5$).}\label{fig4}
\end{figure}


\begin{thebibliography}{}
\bibitem[Archontis et al.(2004)]{archontis2004}Archontis, V., Moreno-Insertis, F., Galsgaard, K., Hood, A., and O'Shea, E., 2004, A\&A, 426, 1047
\bibitem[Bennett et al.(1999)]{bennett1999}Bennett, K., Roberts, B. and Narain, U., 1999, \solphys, 185, 41
\bibitem[Borovsky(2008)]{borovsky2008}Borovsky, J. E., 2008, \jgr, 113, A08110
\bibitem[Brown et al.(2003)]{brown2003}Brown, D. S., Nightingale, R. W., Alexander, D., Schrijver, C. J., Metcalf, T. R., Shine, R. A., Title, A. M., and Wolfson, C.J., 2003, \solphys, 216, 79
\bibitem[Bruno et al.(2001)]{bruno2001}Bruno, R., Carbone, V., Veltri, P., Pietropaolo, E. and
Bavassano, B., 2001, Plan. Space Sci. 49, 1201
\bibitem[Bruno et al.(2007)]{bruno2007}Bruno, R., D'Amicis, R., Bavassano, B., Carbone, V. and
L. Sorriso-Valvo, 2007, Ann. Geophys. 25, 1913
\bibitem[Cartwright and Moldwin(2010)]{cartwright2010}Cartwright, M. L. and Moldwin, M. B., 2010, \jgr, 115, A08102
\bibitem[Chandrasekhar(1961)]{chandrasekhar1961}Chandrasekhar, S., 1961, Hydrodynamic and Hydromagnetic Stability (Oxford: Clarendon Press)
\bibitem[Chang et al.(2004)]{chang2004}Chang, T., Tam, S. and Wu, C., 2004, Phys. Plasmas, 11, 1287
\bibitem[Drazin \& Reid(1981)]{drazin1981}Drazin, P. G., and Reid, W. H., 1981, Hydrodynamic Stability (Cam\-bridge: Cambridge University Press)
\bibitem[Dungey \& Loughhead(1954)]{dungey1954}Dungey, J. W. and Loughhead, R. E., 1954, Austr. J. Phys., 7, 5
\bibitem[Feng et al.(2007)]{feng2007}Feng, H. Q., Wu, D. J., and Chao, J. K., 2007, \jgr, 112, A02102
\bibitem[Feng et al.(2011)]{feng2011}Feng, H. Q., Wu, D. J., Wang, J. M. and Chao, J. W., 2011, A\&A, 527, A67
\bibitem[Lepping et al.(1995)]{lepping1995}Lepping, R. P., Acuna, M. H., Burlaga, L. F. et al., 1995, Space Sci. Rev.
71, 207
\bibitem[Li et al.(2012)]{li2012}Li, X., Morgan, H., Leonard, D. and Jeska, L., 2012, \apjl, 752, L22
\bibitem[Lin et al.(1995)]{lin1995}Lin, R. R., Anderson,  K. A., Ashford, S. et al., 1995, Space Sci. Rev., 71, 125
\bibitem[Lundquist(1951)]{lundquist1951}Lundquist, S., 1951, Phys. Rev. 83, 307
\bibitem[Malara et al.(1996)]{malara1996}Malara, F., Veltri, P., and Carbone, V., 1996, \jgr, 101, 21597
\bibitem[Moldwin et al.(2000)]{moldwin2000}Moldwin, M. B., Ford, S., Lepping, R., Slavin, J., and Szabo, A., Geophys. Res. L., 27, 57
\bibitem[Moreno-Insertis \& Emonet(1996)]{moreno-insertis1996}
Moreno-Insertis, F. and Emonet, T., 1996, \apj, 472, L53
\bibitem[M\"ostl et al.(2009)]{mostl2009}M\"ostl, C., Farrugia, C. J., Biernat, H. K., Leitner, M., Kilpua, E. K. J., Galvin, A. B., Luhmann, J. G., 2009, \solphys, 256, 427
\bibitem[Ogilvie et al.(1995)]{ogilvie1995}Ogilvie, K. W., Chornay,  D. J.,  Fritzenreiter, R. J. et al., 1995, Space Sci. Rev.
71, 55
\bibitem[Osman et al.(2012)]{osman2012}Osman, K., Matthaeus, W., Wan, M. and Rapazzo, A., 2012, Phys. Res. Lett., 108, 261102
\bibitem[Servidio et al.(2009)]{servidio2009}Servidio, S., Matthaeus, W. H.,  Shay, M. A. , Cassak,  P. A.
and Dmitruk, P., 2009, Phys. Rev. Lett., 102, 115003
\bibitem[Srivastava et al.(2010)]{srivastava2010}Srivastava, A. K., Zaqarashvili, T. V., Kumar, P., and Khodachenko, M.L., 2010, \apj, 715, 292
\bibitem[Telloni et al.(2012)]{telloni2012}Telloni, D., Bruno, R., D'Amicis, R., Pietropaolo, E., and
Carbone, V., 2012, \apj, 751, 19
\bibitem[Wedemeyer-B\"ohm et al.(2012)]{wedemeyer2012}Wedemeyer-B\"ohm, S., Scullion, E., Steiner, O., Rouppe van der Voort, L., de La Cruz Rodriguez, J., Fedun, V., and Erd\'elyi, R., 2012, Nature, 486, 505
\bibitem[Zaqarashvili et al.(2010)]{zaqarashvili2010}Zaqarashvili, T. V., D\'iaz, A. J., Oliver, R., and Ballester, J. L., 2010, A\&A, 516, A84
\bibitem[Zaqarashvili et al.(2014)]{zaqarashvili2013}Zaqarashvili, T. V., V\"or\"os, Z. and Zhelyazkov, I., 2014, A\&A, 561, A62
\bibitem[Zhang et al.(2007)]{zhang2007}Zhang, J., Li, L., and Song, Q., 2012, \apj, 662, L35
\bibitem[Zhelyazkov \& Zaqarashvili(2012)]{zhelyazkov2012}Zhelyazkov, I. and Zaqarashvili, T. V., 2012, A\&A, 547, A14
\end{thebibliography}
\end{document}